\documentclass{svjour3}                     
\smartqed  
\usepackage{graphicx,subfigure,amssymb}
\begin{document}

\title{Weak lensing, dark matter and dark energy}


\author{Dragan Huterer}


\institute{Dragan Huterer \at
              Physics Department \\
              University of Michigan\\
              450 Church St.\\
              Ann Arbor, MI 48109\\
              \email{huterer@umich.com}           
}

\date{Received: date / Accepted: date}

\maketitle
 
\begin{abstract}
  Weak gravitational lensing is rapidly becoming one of the principal probes
  of dark matter and dark energy in the universe. In this brief review we
  outline how weak lensing helps determine the structure of dark matter halos,
  measure the expansion rate of the universe, and distinguish between modified
  gravity and dark energy explanations for the  acceleration of the
  universe. We also discuss requirements on the control of systematic
  errors so that the systematics do not appreciably degrade the power of weak
  lensing as a cosmological probe.
\end{abstract}

\section{Introduction}
\label{intro}
Within the past decade, weak gravitational lensing --- slight distortions of
galaxy images due to the bending of the light from distant galaxies by the
intervening large-scale structure --- has become one of the principal probes
of cosmology. The weak lensing regime corresponds to the intervening surface
density of matter being much smaller than some critical value.  While weak
lensing around individual massive halos was measured in the 1990s
\cite{Tyson90,Brainerd95}, weak lensing by large-scale structure was eagerly
expected, its signal predicted by theorists around the same time
\cite{Miralda-Escude91,Kaiser92,Jain_Seljak,Bernardeau_vW_Mellier,Kaiser98,Kamionkowski97,Hui99}.
In this latter regime, the observed galaxies are slightly distorted (roughly
at the 1\% level) and one needs a large sample of foreground galaxies in order
to separate the lensing effect from the noise represented by random
orientations of galaxies.  A watershed moment came in the year 2000 when four
research groups nearly simultaneously announced the first detection of weak
lensing by large-scale structure
\cite{Bacon_detect,Kaiser_detect,vW_detect,Wittman_detect}. Since that time,
weak lensing has grown into an increasingly accurate and powerful probe of
dark matter and dark energy.

The principal power of weak lensing comes from the fact that it responds only
to dark matter, and not to visible (or,  more generally, baryonic)
matter like most other probes of the large-scale structure. Therefore,
modeling  of the visible-to-dark matter bias, a thorny and complicated
subject, is altogether avoided when using weak lensing.  Simulations of dark
matter clustering are becoming increasingly accurate, and in principle there
is no reason why simulation-based predictions cannot reach the 
accuracy required to model the weak lensing signal so that
modeling errors do not appreciably contribute to the total error
budget. Because of our ability to model its signal accurately, 
weak lensing has great intrinsic power to probe dark matter and dark energy in
the universe.

The other principal reason why weak lensing is powerful comes from the nature
of its observable quantity. Galaxy shear is measured relatively
straightforwardly, and comes from millions of galaxies typically observed in
current surveys. While {\it individual} galaxy shear measurements do not
provide cosmological information, correlation of galaxy shear across the sky
does, especially if redshift information of galaxies is utilized to obtain
3-dimensional ``tomography''. In fact, shear-shear correlations can be used to
obtain stringent constraints on cosmological parameters {\it and} use
information from the survey to control the systematic errors at the same time.

In this article, we briefly review the physics of how weak lensing probes dark
matter and dark energy. Our emphasis is on the latter, as dark energy has
emerged as the principal mystery of physics and cosmology (for a review of dark
energy, see \cite{FriTurHut}). The discovery of dark energy and observations
of weak lensing are both about a decade old, and there have been numerous
investigations in recent years as to how best to use weak lensing to probe
dark energy.  This is a condensed review of the subject; for a more extensive
treatment of weak lensing, we recommend excellent reviews by Mellier
\cite{Mellier99}, Bartelmann and Schneider \cite{Bartelmann_Schneider},
Refregier \cite{Refregier_ARAA}, Munshi et al.\ \cite{Munshi_review}, and
Hoekstra and Jain \cite{Hoekstra_Jain}.

\section{Weak lensing basics} \label{sec:basics}

\subsection{Theoretical foundations}

In Newtonian Gauge the perturbed Friedmann-Robertson-Walker metric reads
\begin{equation}
ds^2 = -\left(1+2\Psi\right ) dt^2 + 
 a^2(t)\left (1-2\Phi\right ) 
 \left [d\chi^2 + r^2(d\theta^2+\sin^2\theta d\phi^2) \right  ]
\label{eq:metric}
\end{equation}
where $\chi$ is the radial coordinate, $\theta$ and $\phi$ are angular
coordinates, $a$ is the scale factor, and we have set $c=1$. Here $r(\chi)$ is
the comoving distance; in the special case of a flat universe, $r(\chi)=\chi$.
Finally $\Phi$ and $\Psi$ are the gravitational potentials, and $\Phi\simeq
\Psi$ in General Relativity and in the absence of anisotropic stresses.

 Gravitational lensing produces distortions of images of background
galaxies. These distortions can be described as mapping between the
source plane ($S$) and the image plane ($I$) \cite{Hu_White}

\begin{equation}
\delta x_i^S=A_{ij} \delta x_j^I,
\end{equation}

\noindent where $\delta {\bf x}$ are the displacement vectors in the
two planes and $A$ is the distortion matrix

\begin{equation}
A=
\left ( 
\begin{array}{cc}
	1-\kappa-\gamma_1	&	-\gamma_2   \\
	-\gamma_2		& 1-\kappa+\gamma_1 
\end{array}
 \right ).  
\end{equation}

The deformation is described by the convergence $\kappa$ and
complex shear $\gamma=\gamma_1+i \gamma_2$. We are interested in the
weak lensing limit, where $|\kappa|$, $|\gamma|\ll 1$.  The convergence in
any particular direction on the sky ${\bf\hat n}$ is
given by the integral along the line-of-sight

\begin{equation}
\kappa({\bf \hat n}, \chi)=\int_0^{\chi} W(\chi')\, 
\delta(\chi')\, d\chi',
\label{eq:conv}
\end{equation}

\noindent where $\delta$ is the relative perturbation in matter
density and

\begin{equation}
W(\chi) = {3\over 2}\,\Omega_M\, H_0^2\,g(\chi)\, (1+z)
\end{equation}

\noindent is a weight function, $\Omega_M$ is the matter energy
density relative to critical, $H_0$ is the Hubble constant, and $z$ is the
redshift. Furthermore

\begin{equation}
g(\chi) = r(\chi)\int_{\chi}^{\infty} 
d\chi' n(\chi') {r(\chi'-\chi) \over r(\chi')}
\longrightarrow {r(\chi) r(\chi_s-\chi) \over r(\chi_s)},
\label{eq:g_chi}
\end{equation}

\noindent where $n(\chi)$ is the radial distribution of source galaxies
(normalized so that $\int d\chi\, n(\chi)=1$) and the equality that follows the
arrow holds only if all sources are at a single redshift $z_s$. The
distribution of source galaxies often follows a bell-shaped curve whose peak
redshift depends on the depth of the survey.

The quantity that is most easily determined from observations is shear, which
is directly related to the ellipticity of the observed galaxy (in the weak
lensing limit, shear is approximately equal to the average ellipticity). Shear
is given by

\begin{equation}
\gamma\equiv \gamma_1+i\gamma_2=\frac{1}{2}\left (\psi_{,11}-\psi_{,22}\right
)+i\psi_{,12}	
\end{equation}

\noindent where $\psi$ is the projected Newtonian potential,
$\psi_{,ij}=-(1/2)\int g(\chi)\,(\Psi_{,ij} + \Phi_{,ij})\,d\chi$, and commas denote
derivatives with respect to directions perpendicular to the line
of sight. Unfortunately, this quantity is not easily related to
the distribution of matter in the universe and the cosmological
parameters. Convergence, on the other hand, is given by

\begin{equation}
\kappa=\frac{1}{2}\left (\psi_{,11}+\psi_{,22} \right )
\end{equation}

\noindent which can be directly related to the
distribution of matter (see Eq.~(\ref{eq:conv})), and is convenient for
comparison with theory.  However, it is very difficult to measure the
convergence itself, as convergence depends on the magnification of galaxies
which would need to be measured.

\begin{figure}[!t]
\centerline{\includegraphics[angle=-90,width=6cm]{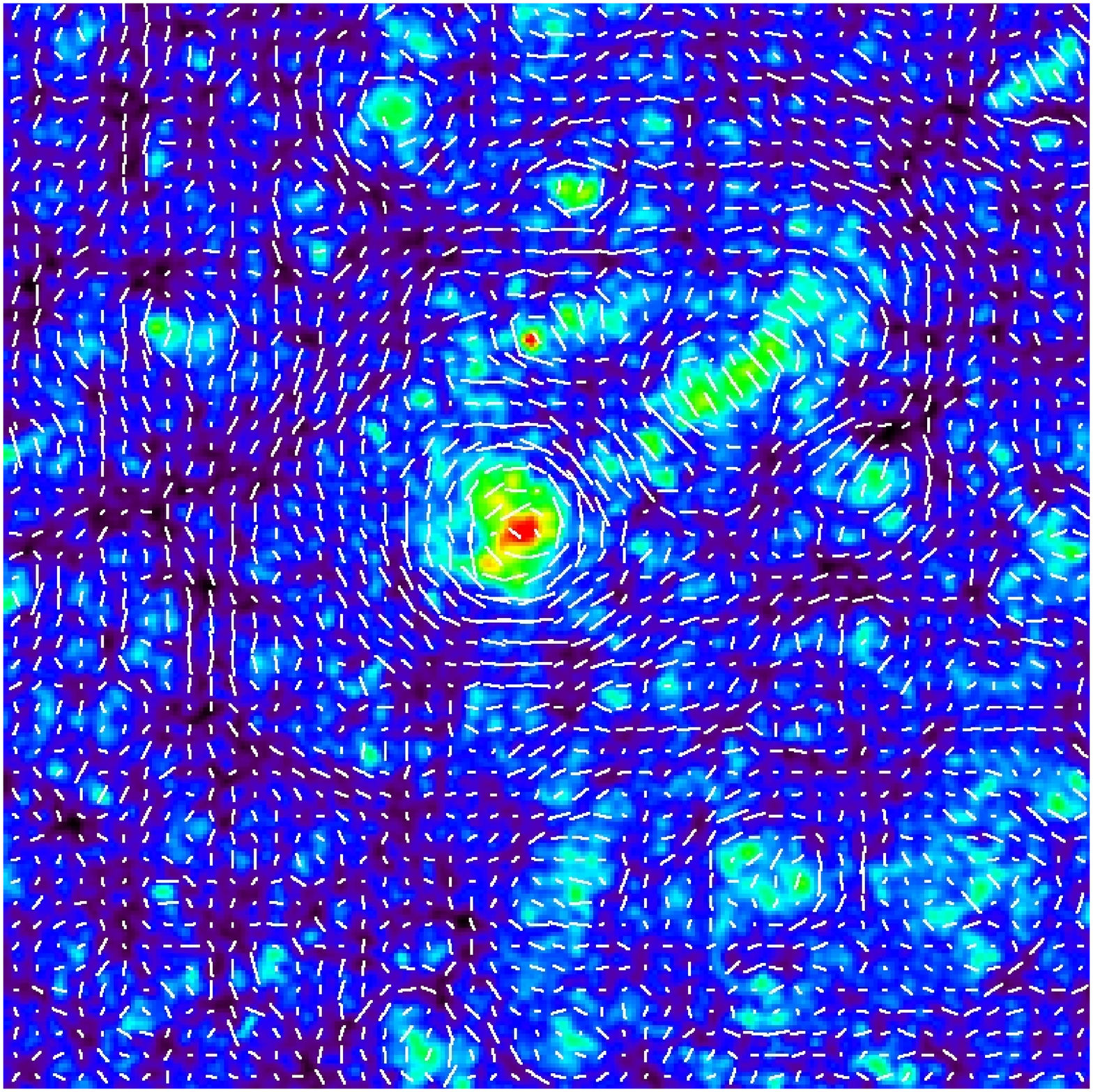}}
\caption{Cosmic shear field (white ticks) superimposed on the 
projected mass distribution from a cosmological N-body simulation: overdense
regions are bright, underdense regions are dark. Note how the shear field is
correlated with the foreground mass distribution.  Figure courtesy of T.\
Hamana.  }
\label{fig:Jain_whiskers}
\end{figure}

Measurements of ellipticities of distant galaxies can be used to reconstruct
the ``reduced shear'' $\gamma/(1-\kappa)$, which is
approximately equal to shear $\gamma$ in the weak lensing limit (for the
potential deleterious effects of assuming this approximation when estimating
cosmological parameters, see \cite{Shapiro_reduced}). Measurements of shear
are complicated by atmospheric effects such as blurring by the atmosphere, the
fact that shapes of galaxies are irregular and not perfectly elliptical,
shearing due to the instrument, etc. Correction of these effects and
minimization of the underlying biases are very important. Recently, impressive
attempts to find agreement between different shear estimators have been made
in the weak lensing community with the Shear TEsting Programme (STEP
\cite{STEP1,STEP2}).

\subsection{Shear/convergence power spectrum and tomography}

While individual galaxy shear (or convergence) cannot be predicted
theoretically, statistical {\it correlations} of galaxy shears can. Transforming the
convergence into multipole space
\begin{equation}
\kappa_{\ell m}=\int d {\bf\hat n} \,\kappa({\bf \hat n}, \chi)\, 
Y_{\ell m}^*,
\end{equation}
and assuming statistical isotropy, the power spectrum of convergence
$P_\ell^{\kappa}$ is then defined as the harmonic transform of the two-point
correlation function of the convergence
\begin{equation}
\langle \kappa_{\ell m}\kappa_{\ell'm'}\rangle =
\delta_{\ell_1 \ell_2}\, \delta_{m_1 m_2}  \,P^{\kappa}(\ell).
\end{equation}

The convergence power spectrum is identical to the shear power spectrum in the
limit of weak distortions; $P^{\gamma}(\ell)\simeq P^{\kappa}(\ell)$. Either
quantity is the observable quantity that contains a lot of information about
the weak lensing field.  In the limit of a Gaussian field, the two point
function would contain all information; however, since the weak lensing field
is nongaussian on small scales, higher-order correlations are important; see
the next subsection.

Weak lensing tomography --- slicing of the shear signal  in  redshift bins --
enables extraction of additional information from the weak lensing shear, as
it makes use of the radial information \cite{Hu_tomo}. Consider correlating
shears in some redshift bin $i$ to those in the redshift bin $j$. The
tomographic cross-power spectrum for these two redshift bins, at a given
multipole $\ell$, is defined by
\begin{equation}
\langle \kappa_{\ell m, i}\kappa_{\ell'm', j}\rangle =
\delta_{\ell_1 \ell_2}\, \delta_{m_1 m_2}  \,P_{ij}^{\kappa}(\ell).
\end{equation}
and can be related to theory via  \cite{Kaiser92,Kaiser98,Hu_tomo}
\begin{equation}
P_{ij}^{\kappa}(\ell) = 
\int_0^{\infty} dz \,{W_i(z)\,W_j(z) \over r(z)^2\,H(z)}\,
 P\! \left ({\ell\over r(z)}, z\right ),
\label{eq:pk_l}
\end{equation} 
\noindent where $r(z)$ is the comoving angular diameter distance and $H(z)$ is
the Hubble parameter.  The weights $W_i$ are given by $W_i(\chi) = {3\over
  2}\,\Omega_M\, H_0^2\,g_i(\chi)\, (1+z)$ where $g_i(\chi) =
r(\chi)\int_{\chi}^{\infty} d\chi_s n_i(\chi_s) r(\chi_s-\chi)/r(\chi_s)$, and
$n_i$ is the comoving density of galaxies if $\chi_s$ falls in the distance
range bounded by the $i$th redshift bin and zero otherwise.
 The observed convergence auto-correlation power spectrum has additional
 contribution from the shot noise given by random galaxy shapes
\begin{equation}
C^\kappa_{ij}(\ell)=P_{ij}^{\kappa}(\ell) + 
\delta_{ij} {\langle \gamma_{\rm int}^2\rangle \over \bar{n}_i}~,
\label{eq:C_obs}
\end{equation}
\noindent where $\langle\gamma_{\rm int}^2\rangle^{1/2}$ is the rms intrinsic
shear in each component (which is typically around $0.2$), and $\bar{n}_i$ is
the average number of galaxies with well-measured shapes in the $i$th redshift
bin per steradian. Clearly, the more galaxies there are and the fewer their
intrinsic ellipticities are, the smaller the shot noise will be.

\begin{figure}[!t]
\centerline{\includegraphics[angle=-90,width=8cm]{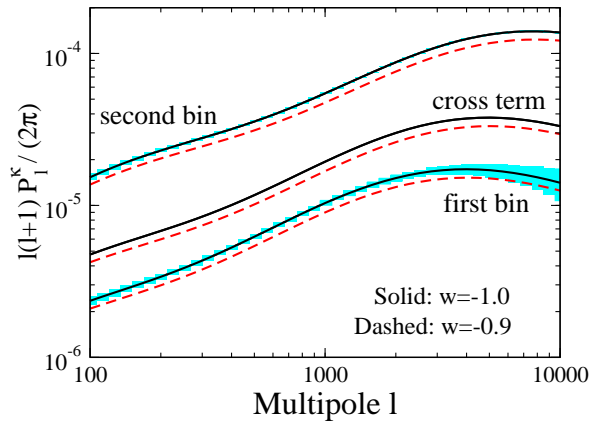}}
\caption{ Cosmic shear angular power spectrum and statistical errors expected
  for LSST for $w=-1$ and $-0.9$. For illustration, results are shown for
  source galaxies in two broad redshift bins, $z_s=0$-$1$ (first bin) and
  $z_s=1$-$3$ (second bin); the cross-power spectrum between the two bins (cross
  term) is shown without the statistical errors.  We employ the redshift
  distribution of galaxies of the form $n(z)\propto z^2\exp(-z/z_0)$ that
  peaks at $2z_0=1.0$. Adopted from Ref.~\cite{FriTurHut}. }
\label{fig:P_kappa_tomo}
\end{figure}

The statistical uncertainty in measuring the shear power spectrum is \cite{Kaiser92,Kaiser98}
\begin{equation}
\Delta P_{ij}^\kappa(\ell) = \sqrt{\frac{2}{(2\ell+1)f_{\rm sky}} } \left[
  P_{ij}^\kappa (\ell) + {\langle \gamma_{\rm int}^2\rangle \over \bar{n}_i}
  \right]~~,
\label{eqn:power_error}
\end{equation}
where $f_{\rm sky}$ is the fraction of sky area covered by the survey. The
first term in brackets, which dominates on large scales, comes from cosmic
variance of the mass distribution, and the second, shot-noise term results
from both the variance in galaxy ellipticities (``shape noise'') and from
shape-measurement errors due to noise in the
images. Fig.~\ref{fig:P_kappa_tomo} shows the dependence on the dark energy of
the shear power spectrum and an indication of the statistical errors expected
for a survey such as Large Synoptic Survey Telescope (LSST \cite{LSST}),
assuming a survey area of 15,000 sq. deg.

\subsection{Three-point correlation function of
  shear/convergence}\label{sec:3pt}

The two-point correlation function has been by far the most analyzed, and so
far best measured, weak lensing statistic. However, higher-order correlation
functions also contain significant cosmological information; chief among these
is the three-point correlation function. With purely Gaussian shear or
convergence field, the three-point function is of course zero (within
measurement and cosmic-variance errors). While primordial nongaussianity can
produce a nonzero three-point function, nonlinear clustering of structure
presents the most widely anticipated --- and in fact guaranteed --- signal in the three-point
function
\cite{Bernardeau_vW_Mellier,Cooray_Hu_bisp,Schneider_Lombardi_I,Schneider_Lombardi_II,TJ_3pt_WL,Dodelson_Zhang_bisp}.

The three-point correlation function of the convergence in multipole space can
be related to the bispectrum of the convergence $B_{\ell_1 \ell_2
  \ell_3}^\kappa$ via
\begin{equation}
\left< \kappa_{\ell_1 m_1} \kappa_{\ell_2 m_2} \kappa_{\ell_3 m_3} \right> 
= 
\left(
\begin{array}{ccc}
 l_1  &  l_2  & l_3 \\
 m_1  &  m_2   &  m_3
 \end{array}
\right)
B_{\ell_1 \ell_2 \ell_3}^\kappa \,
\end{equation}
where the term in parentheses is the Wigner 3j symbol. The convergence
bispectrum is essentially a projection along the line of sight of the matter
overdensity bispectrum
\begin{eqnarray}
B_{\ell_1 \ell_2 \ell_3}^\kappa &=& \sqrt{(2\ell_1+1)(2\ell_2+1)(2\ell_3+1)
  \over 4\pi} 
\left(
 \begin{array}{ccc}
  l_1  &  l_2  & l_3 \\
   0  &  0    &  0
 \end{array}
\right)
        \nonumber\\[0.2cm]
&\times& \left[ \int dz {[W(z)]^3 \over r(z)^4 H(z)}  B\left({\ell_1 \over
        r(z)},{\ell_2 \over r(z)},{\ell_3\over r(z)}, z\right)\right] \, 
\label{eq:bispectrum}
\end{eqnarray}
where $W(z)$ is the same weight function as in Eq.~(\ref{eq:pk_l}). The
(convergence) bispectrum is defined only if the following relations are
satisfied: $|\ell_j-\ell_k|\leq \ell_i \leq |\ell_j+\ell_k|$ for $\{i, j,
k\}\in \{1, 2, 3\}$ and $\ell_1+\ell_2+\ell_3$ is even. To compute the
bispectrum of the convergence, therefore, we need to supply the matter
bispectrum $B(k_1, k_2, k_3, z)$ which, on mildly nonlinear scales of
interest, can be evaluated using the halo model or perturbation theory
techniques, or else calibrated directly from cosmological N-body simulations.

Measurements (and detections) of the three-point function have been first made
in the VIRMOS-DESCART survey \cite{Bernardeau_VIRMOS_3pt,Pen_VIRMOS_skew},
and, more recently, in the COSMOS survey \cite{Semboloni_COSMOS_3pt}.  Much
better constraints are anticipated from future wide-field surveys
\cite{TJ_PS+Bisp}.  However, including the covariance between the power
spectrum and the bispectrum --- a difficult quantity to estimate in its own
right --- may degrade the combined constraints relative to a naive
(i.e.\ uncorrelated) combination of their constraints \cite{Takada_Bridle}.

\section{Galaxy-galaxy lensing and galaxy cluster counts}\label{sec:dm}

\begin{figure}[t]
\centerline{  \includegraphics[width=8cm]{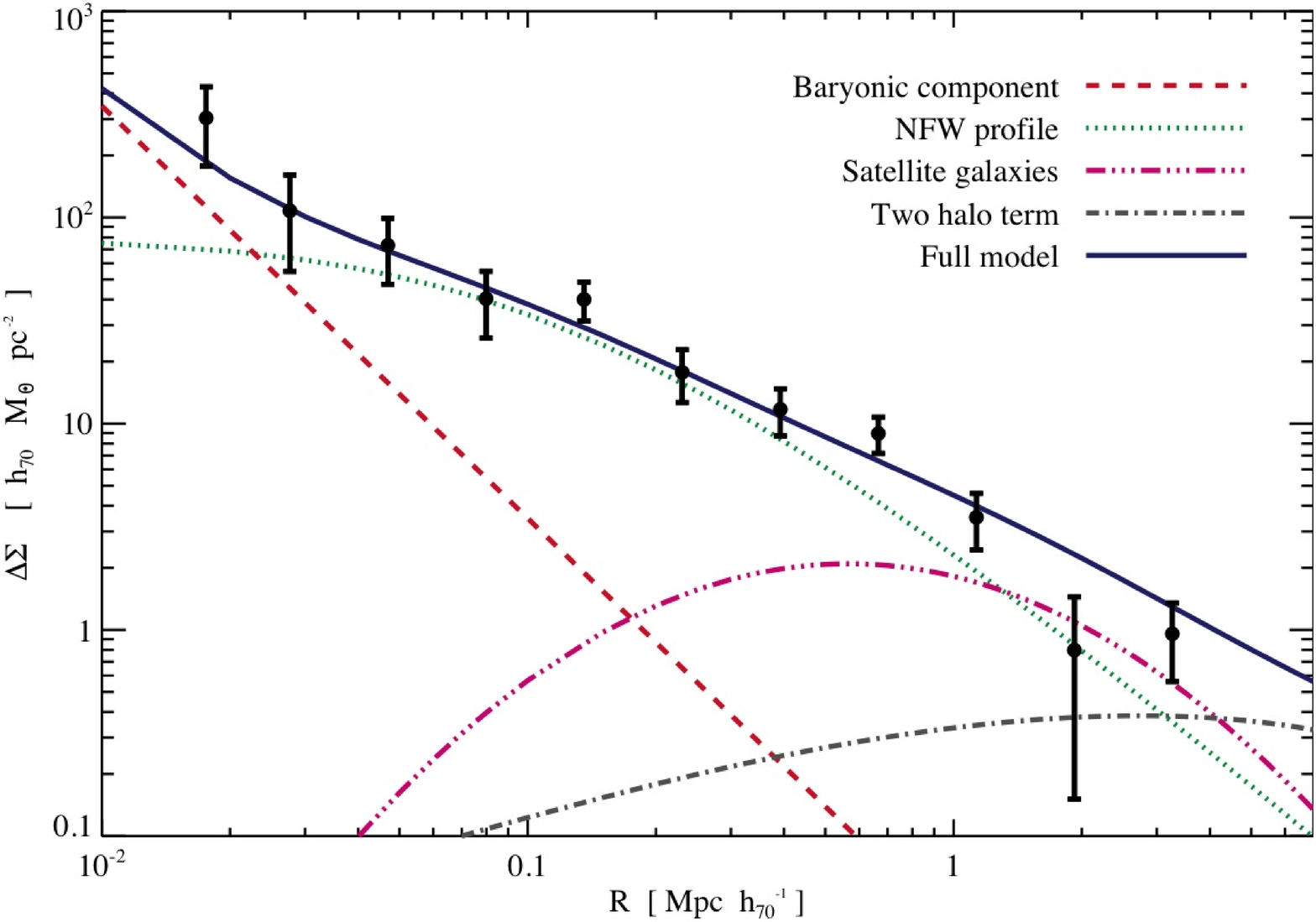}}
\caption{Surface density profile measurements obtained from galaxy groups in
  the COSMOS survey \cite{Leauthaud_measure,Leauthaud_inprep}.  The legend shows
  various components of the halo model fit to the profile. Figure kindly
  provided by Alexie Leauthaud. }
\label{fig:Sigma_R}      
\end{figure}

One effective application of weak lensing is to measure the correlation of
shear of the background galaxies with mass of the foreground galaxies. This
method goes under the name of ``galaxy-galaxy lensing''
\cite{Brainerd95,Fischer,McKay,Hoek_Yee_Glad,Mandelbaum_starform}, and
essentially measures the galaxy-shear correlation function across the sky.
Galaxy-galaxy lensing measures the surface mass density contrast
$\Delta\Sigma(R)$
\begin{equation}
  \Delta \Sigma(R)\equiv\overline{\Sigma}(<R)-\overline{\Sigma}(R)=\Sigma_{\rm crit}\times\gamma_t(R).
\label{dsigma}
\end{equation}
$\overline{\Sigma}(< R)$ is the mean surface density within proper radius $R$,
$\overline{\Sigma}(R)$ is the azimuthally averaged surface density at radius
$R$ (e.g.\ \cite{Miralda-Escude91,Wilson:2001}), $\gamma_t$ is the
tangentially projected shear, and the critical surface density $\Sigma_{\rm
  crit}$ is a known function of the distances to the source and the
lens. The surface density can be further inverted to obtain radial density
profiles of dark matter halos \cite{Sheldon04,Johnston05}. More accurate
modeling of the galaxy-galaxy lensing signal, that takes into account
clustering with other halos and the baryonic content of each halo, can be done
using the halo model of large scale structure \cite{Guzik_Seljak}. Figure
\ref{fig:Sigma_R} shows the surface density profile of a sample of $\sim$600
galaxies in the COSMOS survey with stellar masses of order $\sim 10^{11}
M_{\odot}$ and at $z\sim 0.5$, together with halo model fits
\cite{Leauthaud_measure,Leauthaud_inprep}. Current measurements constrain the
density profiles and bias of dark matter halos
\cite{Sheldon04,Kleinheinrich04,Mandelbaum_profiles,Johnston07}, and the
relation between their masses and luminosities
\cite{Leauthaud_ML,Sheldon09_ML}. In the future, galaxy-shear correlations
have potential to constrain dark energy models \cite{Hu_Jain} and modified
gravity models for the accelerating universe \cite{Schmidt_MGWL}.

\begin{figure}[!t]
\centerline{\includegraphics[width=8cm]{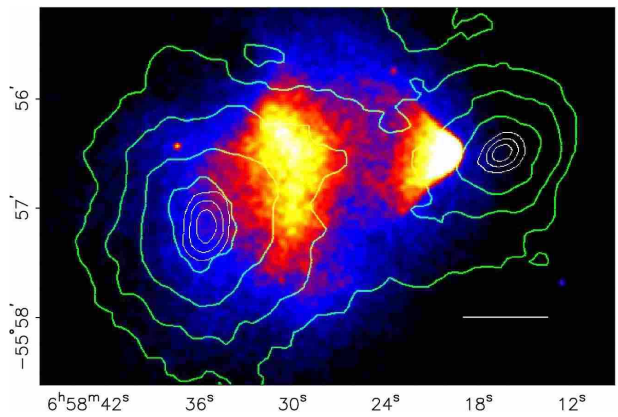}}
\caption{X-ray emission from the ``Bullet'' cluster of galaxies observed by
  ground and space telescopes. Colored features correspond to X-ray emission
  observed by Chandra space telescope, while the green contours correspond to
  mass reconstruction from weak lensing observations. The ``bullet'' is a
  smaller subcluster which has passed through the larger cluster and whose hot
  gas is seen in X-rays, being clearly distinct from the total mass of the
  system.  Adopted from Ref.~\cite{Clowe_Bullet}.  }
\label{fig:bullet_cluster}
\end{figure}

Weak lensing signal can also be used to {\it detect} and count massive halos
--- particularly galaxy clusters. This method, pioneered recently
\cite{Wittman01,Wittman02}, can be used to obtain cluster samples whose masses
are reliably determined, avoiding the arguably more difficult signal-to-mass
conversions required with the X-ray or optical observations
\cite{Wittman06,Schirmer07,Dietrich07,Miyazaki07}.  Much important information
about the dark matter and gas content of galaxy clusters can be inferred with
the combined lensing, X-ray and optical observations. This has recently been
demonstrating with observations of the ``Bullet'' cluster \cite{Clowe_Bullet},
where the dark matter distribution inferred from weak lensing is clearly
offset from the hot gas inferred from the X-ray observations (see
Fig.~\ref{fig:bullet_cluster}), indicating the presence and distinctive
fingerprints of dark matter.

Finally, one can use the abundance clusters detected through weak lensing to
constrain cosmological parameters (e.g.\ \cite{Marian_Bernstein,Fang_Haiman}). The
number of halos of in the mass range $[M, M+dM]$ in a patch
of the sky with solid angle $d\Omega$ and in a redshift interval $[z, z+dz ]$
is given by
\begin{equation}
{d^2 N\over d\Omega dz} = {r^2(z)\over H(z)}\, {dn(M, z)\over dM}\,dM  
\end{equation}
where $r(z)$ and $H(z)$ are the comoving distance and the expansion rate respectively, and
$n(M, z)$ is the number density (the ``mass function'') that can be calibrated from
numerical simulations. Because weak lensing signal is directly sensitive to
the cluster mass, the usual uncertainty in mapping from the observable signal
(X-ray, or optical light, etc) and mass is largely avoided. However,
contamination of the cluster shear by the projected mass (i.e.\ large-scale
structures between us and the cluster) can seriously degrade the effectiveness
of lensing-detected cluster counts (e.g.\ \cite{Hu_Keeton}).  It is therefore necessary to use N-body
simulations to calibrate purity (contribution of false detections) and
completeness (fraction of detections relative to the truth) of the shear peaks
\cite{White_vW_Mackey,Padm_Seljak_Pen,Hamana_Takada_Yoshida,Hennawi_Spergel,Marian_Smith_Bernstein,Dietrich_Hartlap,Kratochvil},
and the jury is still out as to whether the projection systematics can be
controlled to sufficient accuracy. An alternative to using the
weak-lensing-detected clusters is to apply weak lensing mass measurements to
clusters detected via X-ray \cite{Dahle06} or optical \cite{Rozo09}
observations.

\section{Cosmological constraints} \label{sec:cosmoparams}

First detections of weak lensing by large-scale structure have been announced
nearly a decade ago
\cite{Bacon_detect,Kaiser_detect,vW_detect,Wittman_detect}. Their results were
in mutual agreement and consistent with theoretical expectations, which is
remarkable given that they were obtained independently. Meanwhile, additional
measurements have been collected using ground-based
\cite{Jarvis_CTIO,Brown_COMBO17,Hoekstra_RCS,vanWaerbeke_VIRMOS,Hoekstra_CFHT,Hetterscheidt_GaBoDS,Benjamin}
and space-based
\cite{Refregier_HST_MDS,Rhodes_STIS,Leauthaud_measure,Massey_Nature}
observations, leading to increasingly better understanding of systematic
errors.

\subsection{Current constraints}

Current ground-based weak lensing surveys cover of order $\sim 100$ square
degrees.  Space-based surveys currently cover only a few square degrees, but
utilize significantly more distant galaxies whose lensing signal is more
pronounced.

\begin{figure*}[!t]
\centering
\subfigure 
{
    \includegraphics[width=7cm]{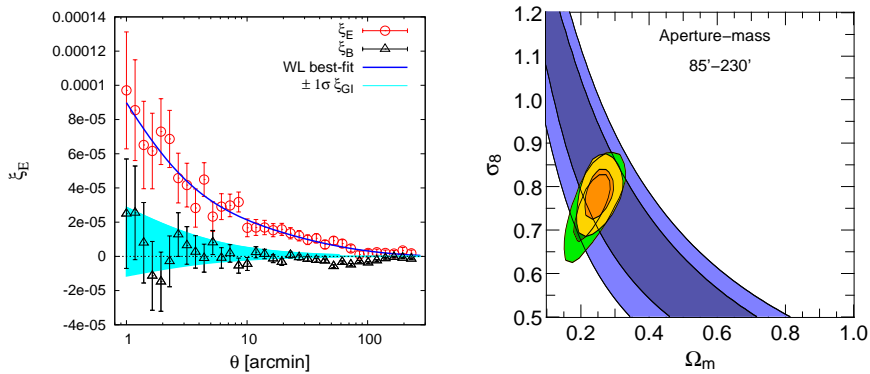}
}
\hspace{-0.4cm}
\subfigure
{
    \includegraphics[width=5cm]{Fu.Om_sig8.ps}
}
\caption{Example of cosmological constraints from weak lensing data, adopted
  from Canada-France-Hawaii Telescope Legacy Survey (CFHTLS;
  \cite{Fu_CFHTLS}). {\it Left panel:} E and B mode measurements; note that B
  modes are consistent with zero, and their scatter is consistent with
  expectations from the alignment between intrinsic shapes of galaxies.  {\it
    Right panel:} Constraints (68\% and 95\% confidence) in the
  $\Omega_M$-$\sigma_8$ plane. Blue contours are from weak lensing, green
  contours are from the CMB (WMAP 3-year; \cite{WMAP3}), and the orange
  contours are the two combined. Adopted from Ref.~\cite{Fu_CFHTLS}.}
\label{fig:Fu_CFHTLS}
\end{figure*}

Modest sky coverage of current surveys, combined with evolving understanding
of the systematic errors, means that only a few parameter combinations can be
measured to an interesting accuracy as of this writing. In particular, the
combination $\sigma_8 \Omega_M^{0.6}$, where $\sigma_8$ is the amplitude of
mass fluctuations, is measured to about 5-10\% from the ground
\cite{Kilbinger_CFHTLS,Benjamin} and $\sim$15\% from space
\cite{Massey_COSMOS}, and is in concordance with values of $\Omega_M\simeq
0.25$ and $\sigma_8\simeq 0.8$. An example adopted from Ref.~\cite{Fu_CFHTLS}
is shown in Fig.~\ref{fig:Fu_CFHTLS}, with measurements of E and B mode
correlation functions of shear in real space (left panel), and the resulting
constraints on $\sigma_8$ and $\Omega_M$ (right panel).  Such constraints are
impressive given that first detections of weak lensing by large-scale
structure have been made less than a decade ago.

Weak lensing also probes dark energy, though at the present time it needs to
be combined with other cosmological probes in order to produce interesting
constraints on the equation of state $w$. Recent work illustrates how weak
lensing helps tighten the constraints from SNe Ia and CMB \cite{Jarvis_CTIO_DE}
mostly due to its sensitivity to growth of density perturbations. It is also
possible to place constraints on specific dark energy model from current weak
lensing data \cite{Schimd_current_Q}. Upcoming weak lensing surveys from the
ground and space hold promise to measure dark energy significantly better, as
we now discuss. 

\subsection{Future constraints}

The errors in the cosmological constraints can then be easily and
straightforwardly forecasted using the Fisher matrix formalism. Consider a set
of cosmological parameters $p_i$ ($i=1,\ldots, N_{\rm par}$) and 
weak lensing observations that result in measurements of the shear power
spectrum $C^{\kappa}_{jk}(\ell)$, defined in Eq.~(\ref{eq:C_obs}) and where
$j$ and $k$ run over the tomographic redshift bins. Then the Fisher matrix is
defined as
\begin{equation}
F_{ij} = \sum_{\ell} \,{\partial {\bf C}\over \partial p_i}\,
{\bf Cov}^{-1}\,
{\partial {\bf C}\over \partial p_j},\label{eq:latter_F}
\end{equation}
\noindent  where ${\bf C}$ is the column matrix of the observed power spectra and
${\bf Cov}^{-1}$ is the inverse of the covariance matrix 
between the observed power spectra whose elements are given by
\begin{equation}
{\rm Cov}\left [C^{\kappa}_{ij}(\ell), C^{\kappa}_{kl}(\ell')\right ] = 
{\delta_{\ell \ell'}\over (2\ell+1)\,f_{\rm sky}}\,
\left [ C^{\kappa}_{ik}(\ell) C^{\kappa}_{jl}(\ell) + 
  C^{\kappa}_{il}(\ell) C^{\kappa}_{jk}(\ell)\right ].
\label{eq:Cov}
\end{equation}
assuming that the convergence field is Gaussian, which is a good approximation
at $\ell\lesssim  3000$.  By the Kramer-Rao inequality, cosmological parameters can
typically be measured to no better than $\sigma(p_i)=1/\sqrt{F_{ii}}$
(unmarginalized error), or $\sigma(p_i)=\sqrt{(F^{-1})_{ii}}$ (marginalized over all
other parameters). In practice, these idealistic error bars are usually a good
approximation to the truth if the parameters are sufficiently well measured so
that their
covariances are ellipsoid-shaped and not ``banana-shaped'' regions.

\begin{figure*}[!t]
\centering
\subfigure 
{
    \includegraphics[angle=-90,width=6cm]{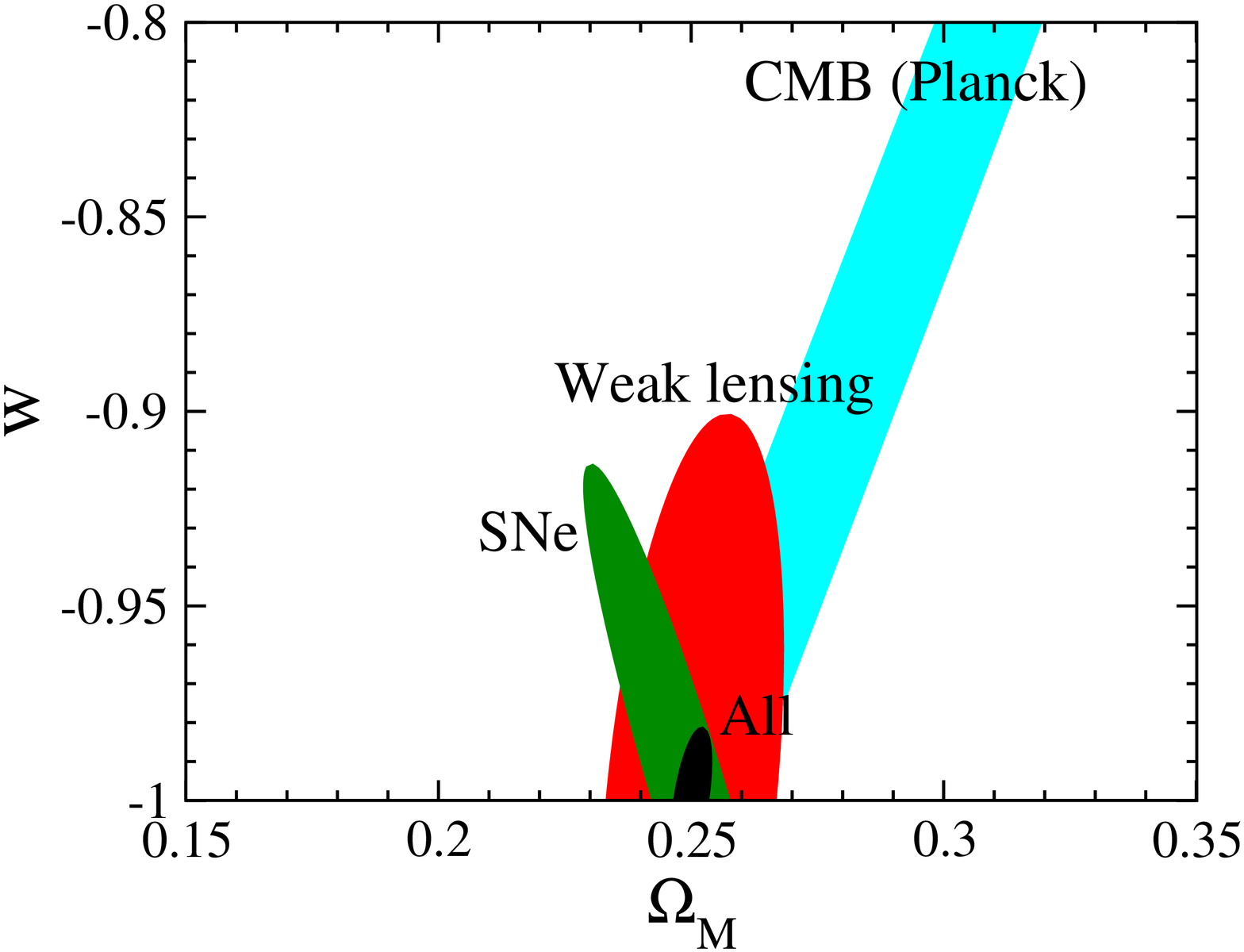}
}
\hspace{-0.4cm}
\subfigure
{
    \includegraphics[angle=-90,width=6cm]{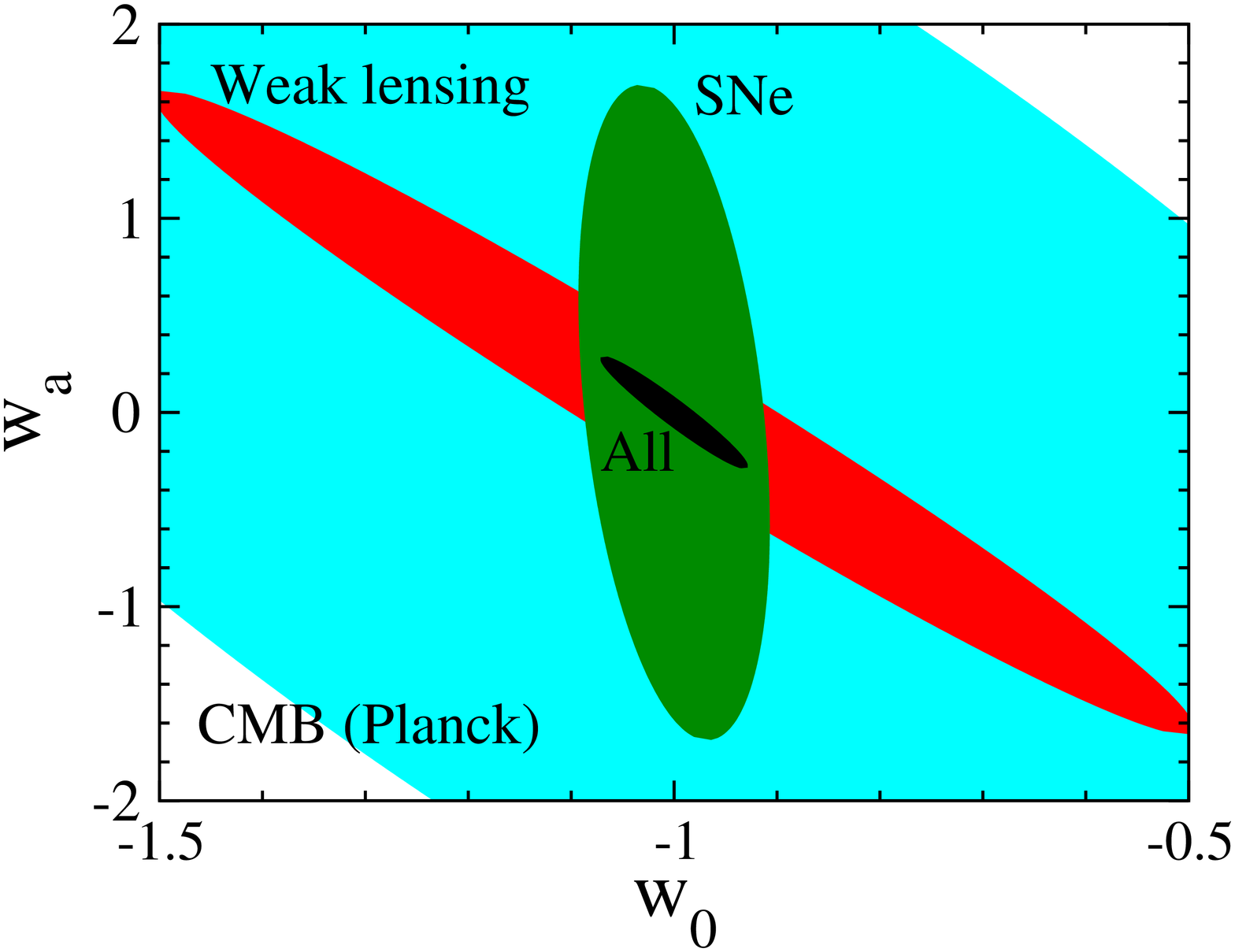}
}
\caption{Illustration of forecast constraints on dark energy parameters,
  adopted from Ref.~\cite{FriTurHut}.  Shown are 68\% C.L.\ uncertainties for
  one version of the proposed SNAP experiment, which combines a narrow-area
  survey of 2000 SNe to $z=1.7$ and a weak lensing survey of 1000 sq. deg.
  {\it Left panel:} Constraints in the $\Omega_{\rm M}$-$w$ plane, assuming
  constant $w$; the vertical axis can also be interpreted as the pivot value
  $w_p$ for a time-varying equation of state. {\it Right panel:} Constraints
  in the $w_0$-$w_a$ plane for time-varying dark energy equation of state,
  marginalized over $\Omega_{\rm M}$ for a flat Universe.  }
\label{fig:Omw_and_w0wa}
\end{figure*}

Weak lensing measurements can determine cosmological parameters to an
excellent accuracy due to the fact that modeling the weak lensing signal only
requires modeling of dark matter, though baryonic collapse at centers of dark
matter halos and galaxy alignments complicate the picture somewhat and
contribute to systematic errors discussed below.

First cosmological parameter accuracy projections from weak lensing have been
made in \cite{Hu_Tegmark_WL} while increasingly more sophisticated estimates
\cite{Huterer_thesis,Simon_King_Schneider,TJ_PS+Bisp,Munshi_Valageas,Heavens_Kitching_Taylor,Amara_Refregier,Zhan_Knox,Shapiro_Dodelson_WL_counts,Bernstein_comprehensive}
indicate prospects for determining the parameter $\sigma_8$ to much better
than 1\% and the equation of state of dark energy to a few percent accuracy in
the future. An example of such projections is shown in
Fig.~\ref{fig:Omw_and_w0wa}, where we show constraints on $\Omega_M$ and
constant $w$ (left panel) and $w_0$ and $w_a$ (right panel) from
SuperNova/Acceleration probe (SNAP \cite{SNAP}), a proposed space-based experiment to
probe dark energy. These constraints marginalize over $\sim 5$ other
cosmological parameters, and allow for the presence of systematic as well as
statistical errors. 

Finally, we mention the proposed technique to correlate the mass of foreground
galaxies with the shear of the background galaxy population, and thus isolate
geometric distance factors in photometric redshift bins. This approach
sometimes goes under the name of ``cross-correlation cosmography''
\cite{Bernstein_Jain_cosmography}, and is very closely related to
galaxy-galaxy lensing.  Cross-correlation cosmography is potentially less
sensitive to systematic errors as well as assumptions about the nonlinear
clustering of dark matter than the shear-shear correlation measurements, and
can significantly complement cosmological constraints obtained from the
latter.
\cite{Bernstein_Jain_cosmography,Jain_Taylor,Zhang_Hui_Stebbins,Bernstein_metric,Taylor_shear_ratio,Hu_Jain}

\section{Weak Lensing and Modified Gravity} \label{MG}

Weak lensing is a particularly effective discriminator of  modified
gravity explanations for the accelerating universe. Modified gravity
stipulates that the reason for the apparent acceleration are modifications to
General Relativity on large scales (e.g.\ in the infrared). General Relativity
has to be preserved on  the solar-system and galactic scales in order to satisfy
stringent observational constraints, but modifications on large scales could
masquerade as an {\it apparent} acceleration when the equations of standard
Einstein's gravity are used to interpret the observations.

\begin{figure*}[!t]
\centering
\subfigure 
{
    \includegraphics[angle=-90,width=6cm]{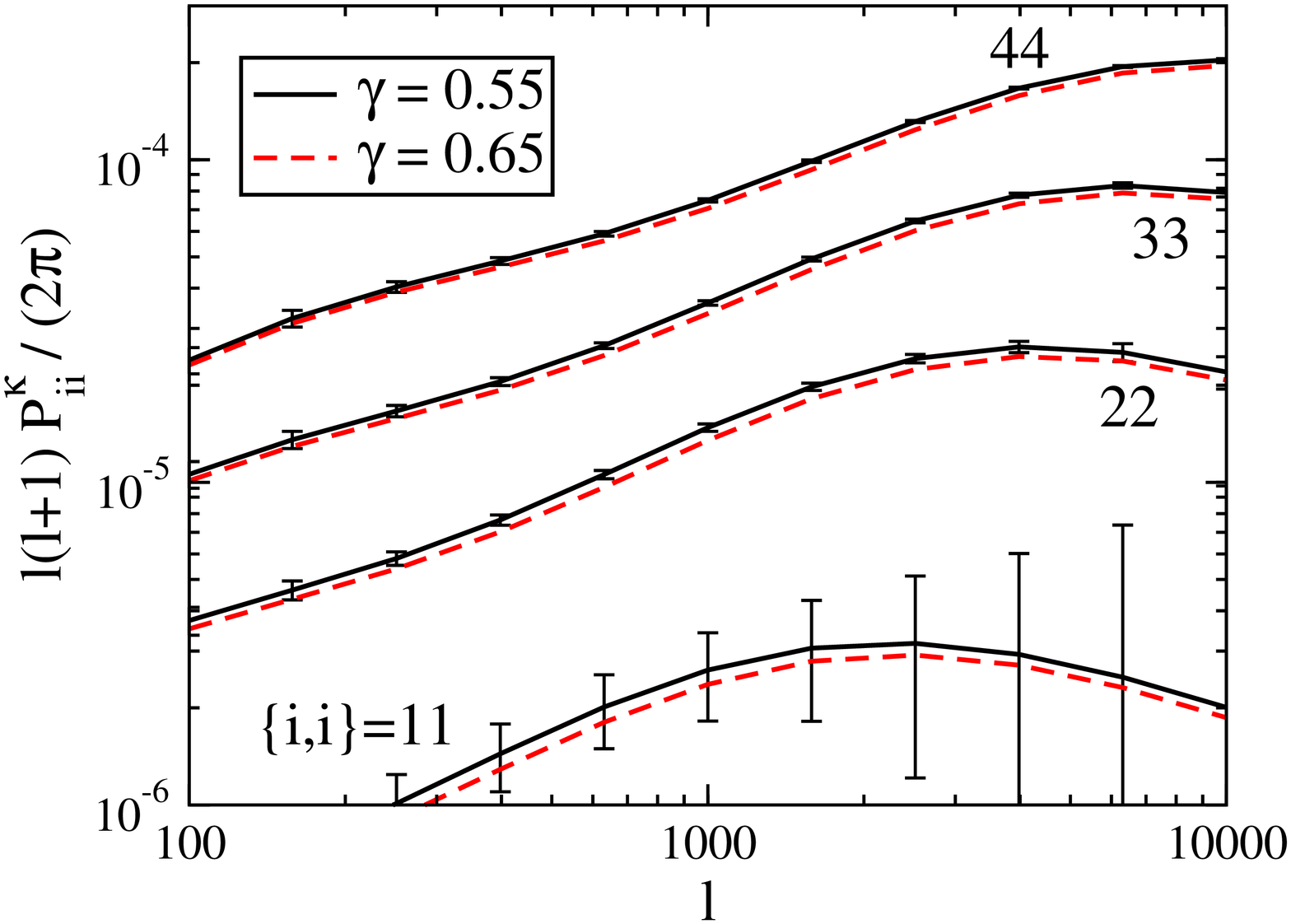}
}
\hspace{-0.4cm}
\subfigure
{
    \includegraphics[angle=-90,width=6cm]{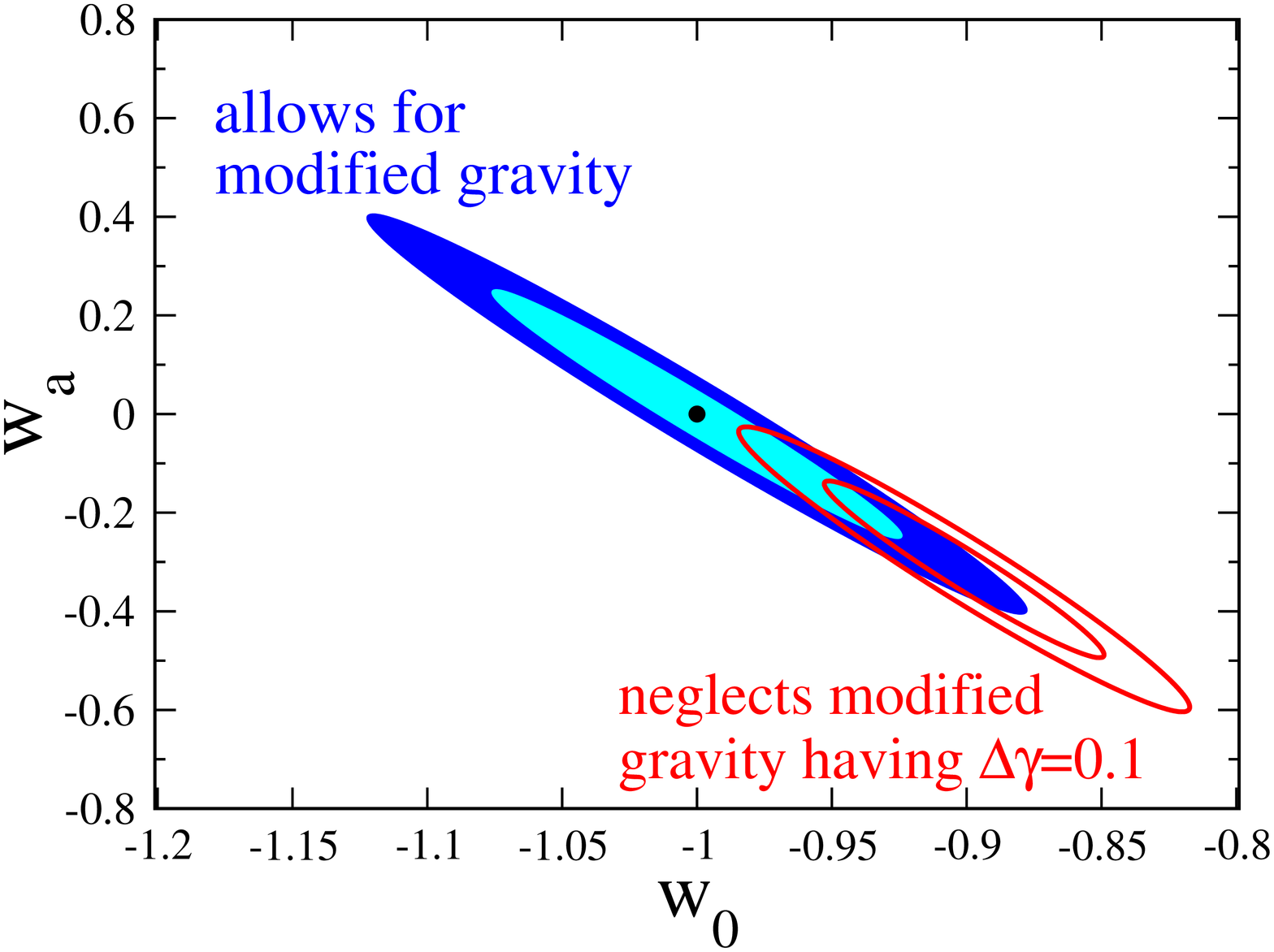}
}
\caption{Example of how weak lensing can constrain modifications of gravity
  parametrized by the growth index $\gamma$, adopted from
  Ref.~\cite{MMG}. {\it Left panel:} Auto-correlation power spectra of 4-bin
  weak lensing tomography for the two values of the growth index, with
  statistical errors shown for the $\gamma=0.55$ model. This shows that a
  relatively small signal due to modified gravity can be detected using future
  weak lensing observations.  {\it Right panel:} Bias in constraints in
  cosmological parameters $w_0$ and $w_a$ if changes in modified gravity were
  ignored, for the $\Delta\gamma=0.1$ modification relative to the fiducial
  model. Recall, the growth index in DGP braneworld gravity differs from the
  General Relativity value by $\Delta\gamma\approx 0.67-0.55=0.12$.  }
\label{fig:gamma}
\end{figure*}

Gravitational lensing is particularly useful probe of modified gravity because
it probes the sum of the two gravitational potentials $\Phi+\Psi$ (see
Eq.~(\ref{eq:metric})), while particle dynamics probes $\Psi$ alone
(e.g.\ \cite{Uzan_Bernardeau,Jain_Zhang}).  Since modifications to gravity typically affect
the potentials differently, combination of weak lensing with other
cosmological probes can in principle be used to differentiate modified gravity
from dark energy. For a detailed review of how weak lensing can be used to
test gravity, see the review \cite{Uzan_GRG} in this Special Issue.

The simplest approach to distinguish between modified gravity and
honest-to-god dark energy is to parametrize anomalous growth of density
perturbations; the most popular approach is via the ``growth index'' parameter
\cite{Linder_gamma}. The linear growth factor $g(a)\equiv \delta(a)/a)$
(scaling out the matter dominated universe behavior $\delta\propto a$),
where $a$ is the scale factor, is approximated by
\begin{equation} 
g(a)= e^{\int_0^a d\ln a\,[\Omega_M(a)^\gamma-1]}, 
\label{eq:groexp}
\end{equation}
where $\gamma$ is a new parameter called the growth index (not to be confused
with shear!).  This fitting function is accurate to 0.3\% compared to the exact
solution within general relativity for a wide variety of physical dark energy
equation of state ratios including equations of state $w(a)=w_0+w_a(1-a)$,
provided that the function takes the value $\gamma = 0.55 + 0.05(1+w(z=1))$
\cite{Linder_gamma}. The fitting formula also fits \cite{LinCah07} the linear
growth in the popular ``braneworld gravity'' model of Dvali, Gabadadze and
Porrati (DGP \cite{DGP}), but with a different value of the growth index,
$\gamma\simeq 0.67$. More generally, measuring $\gamma$ can distinguish between
modified gravity and dark energy even if the two predict identical expansion
history (e.g.\ distances in the universe; for a similar approach not using the
growth index, see \cite{IshUpaSpe,Daniel}). Future expectations are for
$\gamma$ to be measured to accuracy of about 0.02-0.03 from weak lensing data
combined with other cosmological probes that measure the distance scale
(SNe Ia, CMB) or growth (cluster abundance) \cite{MMG,Amend_Kunz_Sapone}; see
Fig.~\ref{fig:gamma} for an example.
 
Weak lensing can also probe specific models where dark matter or dark energy
interactions are modified. For example, models of massive gravity, where the
Poisson equation is phenomenologically modified with the Yukawa term, have
been constrained with weak lensing observations
\cite{Dore_CFHTLS,White_Kochanek}; preliminary constraints have been imposed
on some other models as well \cite{Reyes:2010tr}. In the future, weak lensing
data will enable a much more ambitious battery of tests of modified gravity,
essentially enabling comparison of distance and growth functions at multiple
redshifts and on multiple angular scales
\cite{Schmidt08,Zhan_Knox_Tyson,Zhang_Bean_etal,Zhao_etal_08}.

\section{Second-order effects as a signal for cosmology} \label{sec:second_order}

So far we have discussed weak lensing of galaxies by the large-scale
structure, a principal signature of weak lensing used in cosmology. However,
objects other than galaxies are also weakly lensed, and their properties can
sometimes be used to our advantage.

In particular, high-z type Ia supernovae (SN Ia) will be gravitationally
lensed by the foreground large-scale structure. The useful property of SN Ia
is that they have nearly uniform intrinsic luminosities. Weak lensing will
magnify SN Ia and perturb these luminosities, and the resulting magnifications
will be correlated and may be measurable with future surveys. The resulting
magnification of supernova luminosity distance is
\begin{equation}
\mu = [(1-\kappa)^2-|\gamma|^2]^{-1} \approx 1 + 2 \kappa + 3\kappa^2+|\gamma|^2+...\, ,
\label{eq:higher_order}
\end{equation}
where $\kappa$ is the lensing convergence and
$|\gamma|=\sqrt{\gamma_1^2+\gamma_2^2}$ is the total shear at the angular
location of the SN. Each supernova's luminosity $L$ and distance $d_L$ will be
perturbed as $\delta L/L = 2|\delta d_L/d_L| = \mu-1$, where $\mu$ is
magnification at that location and out to redshift of the SN. The magnification
of SN --- really, the correlated pattern of the departures of their observed
distances from the true values --- comes for free with standard measurements
of SN distances to measure the expansion history of the universe.

Measurements of cosmic magnification of supernovae therefore complement galaxy
shear measurements in providing a direct measure of clustering of the dark
matter \cite{CooHolHut_SNmag}.  As the number of supernovae is typically much
smaller than the number of galaxies, the two-point correlation function of lensed SN Ia is
more noisy than, and not as effective as, the galaxy shear. However, weak lensing of
SN Ia probes magnification directly, and can be used to measure the second-order
corrections in Eq.~(\ref{eq:higher_order}) between magnification and
convergence (or equivalently, shear and convergence)
\cite{CooHolHut_SNmag}. Additionally, the power spectrum of magnification can
be used to measure the cosmological parameters such as $\sigma_8$
\cite{Dodelson_Vallinotto}. While this application of weak lensing is
rather challenging and requires a large number of SN Ia to be effective, we
emphasize that it comes for free in wide-field surveys that measure galaxy
shear as well as SN Ia, and combines weak lensing shear and magnification
information in the same field, providing a number of cross checks on the
systematics.

\section{Systematic errors and  their calibration requirements} \label{sec:syserr}

Because weak lensing has so much statistical power, it is imperative to
control the systematic errors so that they do not appreciably degrade the
statistical errors on the weak lensing power spectrum (see e.g.\ the small
error bars in Fig.~\ref{fig:P_kappa_tomo}), and the similarly small errors on
the cosmological parameters (see Fig.~\ref{fig:Omw_and_w0wa}).

\subsection{Observational errors}

One of the most important systematic effects on weak lensing are the
photometric redshift errors --- uncertainties in measuring redshifts of
sheared galaxies. For weak lensing measurements that simply integrate signal
along the line of sight, redshift errors are clearly not critically important,
as only the overall radial distribution of galaxies, $n(z)$, needs to be known
(see Eq.~\ref{eq:g_chi}). However, for tomographic measurements, the
photometric redshift errors lead to leakage of galaxies from their original
redshift bin and contamination of signal in the bin where they are misplaced.

\begin{figure}[!t]
\centerline{\includegraphics[width=8cm]{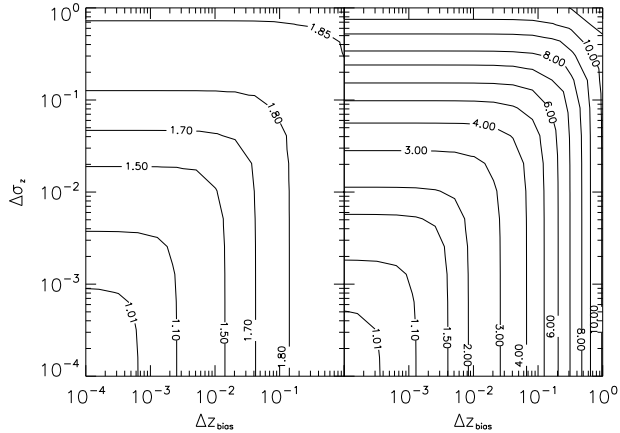}}
\caption{Requirements on the accuracy of photometric redshifts to perform weak
  lensing tomography, adopted from Ref.~\cite{MaHuHut}. Contours show error
  degradations in constant $w$ (left panel) and those in $w_a$ (when where
  both $w_0$ and $w_a$ are varied; right panel), as a function of prior
  knowledge of the mean (x-axis) and scatter (y-axis) in the relation between
  the true (spectroscopic) and photometric redshift. The degradations are
  defined as actual errors in cosmological parameters relative to the errors
  with perfectly known photometric redshifts. For example, the right panel
  shows that no more than 50\% degradation (factor 1.5 on the contours) in the
  error in $w_a$ requires knowledge of the photometric redshift mean and
  scatter to better than $\sim 0.003$ in each redshift bin. }
\label{fig:syserr} 
\end{figure}

The required accuracy of photometric redshift bins has been studied in detail
\cite{MaHuHut}; and it was found that the mean and spread of the relation
between photometric and (true) spectroscopic redshift, {\it averaged} over bins of
$\Delta z=0.1$, both need to be {\it known} to better than about 0.003
\cite{MaHuHut}; see the right panel of Fig.~\ref{fig:syserr}. While errors in
individual galaxy redshift are greater than this by about an order of
magnitude, it is reasonable to expect that modern photometric redshift
techniques (e.g.\ \cite{Oyaizu07}) can achieve this required accuracy in each
redshift bin. Moreover, a possibility of {\it catastrophic}
redshift errors --- cases where the true redshift is misestimated by a
significant amount --- is equally troubling, and requires knowledge of outlier
``islands'' in the $z_{\rm spec}$-$z_{\rm phot}$ plane to a similar
sub-percent accuracy \cite{catastrophic,Hearin_cata}.

Large size of current and future surveys, with millions and up to a billion of
observed galaxies, {\it requires} the use of photometric redshifts. These
photometric redshifts will be calibrated by obtaining spectra of a subset of
galaxies, and ``training'' the photometric spectra using the spectroscopic
subset. Of order $10^5$ spectra seem to be required for future surveys
\cite{MaHuHut,Amara_Refregier,Bernstein_Ma,catastrophic}, although this number strongly depends on the
expected scatter in the photometric redshifts around the true value and other
details. Moreover, the spectroscopic galaxies need to be a representative
sample of the whole population --- for example, very high redshift galaxies in the
survey need to be represented in the spectroscopic subsample --- and this adds
another layer of challenge to the program of photometric redshift calibration.

Errors on the observed shapes of galaxies are manifold, and include imperfect
knowledge of the point-spread function, atmospheric blurring, thermal and
mechanical effects of the telescope, etc.  More generally, errors in
measurements of galaxy shear can be described as either additive or multiplicative 
errors \cite{HTBJ}. The measured shear of source galaxies at redshift $z_s$
and in direction ${\bf n}$, 
$\hat{\gamma}(z_s, {\bf n})$, can be written as 
\begin{equation}
\hat{\gamma}(z_s, {\bf n}) = \gamma\left (z_s, {\bf n}\right )\times 
    [1+f_i(z_s, {\bf n})] +\gamma_{\rm add} (z_s, {\bf n})
\end{equation}
where $\gamma\left (z_s, {\bf n}\right )$ is the true shear, $f_i(z_s, {\bf
  n})$ is the multiplicative and $\gamma_{\rm add}(z_s, {\bf n})$ the additive error
(\cite{HTBJ}; see also \cite{Kitching_sys}).  Multiplicative errors can be
caused by, for example, finite size of the point-spread function with which
the galaxy image is convolved while additive errors can be created, for
example, by atmospheric blurring. The requirements are stringent; for example,
future surveys need to calibrate the mean multiplicative errors to about 0.1\%
(i.e.\ $\delta f_i\approx 0.001$) accuracy in each of $\sim 10$ redshift bins in
order that dark energy constraints not be significantly degraded
\cite{HTBJ}. However, what really helps weak lensing is the possibility of
``self-calibrating'' the systematic errors --- determining a reasonable set of
the systematic error nuisance parameters from the survey concurrently with the
cosmological parameters without appreciable degradation in accuracy on the
latter \cite{HTBJ,Zhang_selfcal,Zhang_Pen_Bernstein}. With self-calibration,
the survey itself is used to partially calibrate the systematic effects. 

\subsection{Theoretical modeling systematics}

Predicting the weak lensing shear power spectrum requires knowledge of the
dark matter power spectrum (see Eq.~\ref{eq:pk_l}); intrinsic uncertainty in
predicting the power spectrum {\it in the nonlinear regime} (on scales
$\ell\gtrsim 200$, or below about 1 degree) constitutes the principal
theoretical weak lensing systematic.  Ref.~\cite{Huterer_Takada} found that
power spectrum on scales of $\sim 1$ Mpc (corresponding roughly to a few
arcminutes on the sky), where weak lensing measurements are most accurate,
needs to be calibrated to a few percent. While not trivial, this accuracy
seems within reach due to recent impressive progress in obtaining precision
results from large numerical simulations (e.g.\ \cite{Heitmann_robustness}).

One additional difficulty, currently only partially explored, is how the
addition of gas to pure dark matter simulations affects the results; it
appears that baryonic physics can be phenomenologically modeled with
sufficient accuracy
\cite{Rudd_Zentner_Kravtsov,Zentner_Rudd_Hu,Hearin_Zentner}. Another challenge
to theoretical predictions is that a large suite of numerical simulations
appears necessary to cover the multi-dimensional cosmological parameter space;
fortunately, clever statistical techniques can be employed to cover
the parameter space so that a very reasonable total number of simulations ($\sim 100$) is sufficient
\cite{Heitmann_calib,Heitmann_CoyoteII}.

Weak lensing provides a large amount of information, given by shapes and
redshifts of millions (or, in the future, billions) of galaxies. Therefore,
there is redundancy in weak lensing measurements that can be used to minimize
the effect of systematic errors with negligible degradation in the statistical
errors on cosmological parameters. For example, by taking only the {\it
  cross}-power spectra for cosmology ($P_{ij}^{\kappa}(\ell)$ with $i\neq j$
in Eq.~(\ref{eq:pk_l})), one can minimize biases due to correlations between
intrinsic galaxy ellipticities \cite{Takada_White} (see also
\cite{Joachimi_Schneider_nulling}). Similarly, one can perform ``nulling
tomography'' of weak lensing to selectively throw out the small-scale
information in the convergence power spectrum that is most sensitive to the
unwanted biases \cite{nulling}.

\section{Conclusions}\label{sec:concl}

Weak lensing probes dark matter and energy in the universe in multiple ways. With
galaxy-galaxy lensing, it probes the structure of individual dark matter
halos.  With detecting galaxy clusters from the shear of the foreground
galaxies, it enables the cosmological number-count test that is sensitive to
distances and growth of density perturbations, and thus to temporal evolution
of dark energy. And with measuring the correlation of shear signal across the
sky, weak lensing again probes distances and growth, but now with a completely
different set of statistical and systematic errors.

In this review article, we briefly covered the aforementioned cosmological
uses of weak lensing; we did not mention some related topics that are covered
elsewhere in this Special Issue (for example, lensing of the CMB
\cite{Hanson_GRG}). We also emphasized the intrinsic power of weak lensing,
and the potential of weak lensing to self-calibrate an impressive range of
systematic errors and still produce excellent cosmological constraints.  This
situation can perhaps be contrasted with strong gravitational lensing, where
only a modest number of multiply imaged systems ($\sim 100$) is currently
available, and where it is important to have very precise {\it a-priori}
constraints on the systematic effects in order to extract the cosmologically
interesting information.

Current efforts in weak lensing research emphasize performing observations
with ever wider and deeper surveys, improving the accuracy and reliability of
weak lensing measurements, improving the understanding of nonlinear clustering
of dark matter and baryons with N-body simulations, and devising clever
methods to self-calibrate the systematic errors and measure new properties of
dark energy and modified gravity. If these efforts continue at the current
rate of progress, weak lensing is likely to firmly establish itself as one of
the most reliable and powerful probes of dark energy and matter in the
universe.

\begin{acknowledgements}
  I would like to thank J\"{o}rg Dietrich and Alexie Leauthaud for useful
  comments on the earlier version of Sec.~\ref{sec:dm}, and Alexie Leauthaud
  for kindly providing Fig.~\ref{fig:Sigma_R}. My work is supported by the DOE
  OJI grant under contract DE-FG02-95ER40899, NSF under contract AST-0807564,
  and NASA under contract NNX09AC89G.
\end{acknowledgements}

\bibliographystyle{spphys}       
\bibliography{wl}   

\begin{thebibliography}{100}
\providecommand{\url}[1]{{#1}}
\providecommand{\urlprefix}{URL }
\expandafter\ifx\csname urlstyle\endcsname\relax
  \providecommand{\doi}[1]{DOI \discretionary{}{}{}#1}\else
  \providecommand{\doi}{DOI \discretionary{}{}{}\begingroup
  \urlstyle{rm}\Url}\fi

\bibitem{Tyson90}
J.A. {Tyson}, R.A. {Wenk}, F.~{Valdes}, Astrophys. J. Lett. \textbf{349}, L1
  (1990)

\bibitem{Brainerd95}
T.G. Brainerd, R.D. Blandford, I.~Smail, Astrophys. J. \textbf{466}, 623 (1996)

\bibitem{Miralda-Escude91}
J.~{Miralda-Escude}, Astrophys. J. \textbf{380}, 1 (1991)

\bibitem{Kaiser92}
N.~Kaiser, Astrophys. J. \textbf{388}, 272 (1992)

\bibitem{Jain_Seljak}
B.~Jain, U.~Seljak, Astrophys. J. \textbf{484}, 560 (1997)

\bibitem{Bernardeau_vW_Mellier}
F.~Bernardeau, L.~Van~Waerbeke, Y.~Mellier, Astron. Astrophys. \textbf{322}, 1
  (1997)

\bibitem{Kaiser98}
N.~Kaiser, Astrophys. J. \textbf{498}, 26 (1998)

\bibitem{Kamionkowski97}
M.~Kamionkowski, A.~Babul, C.M. Cress, A.~Refregier, Mon. Not. Roy. Astron.
  Soc. \textbf{301}, 1064 (1998)

\bibitem{Hui99}
L.~Hui, Astrophys. J. \textbf{519}, L9 (1999)

\bibitem{Bacon_detect}
D.J. Bacon, A.R. Refregier, R.S. Ellis, Mon. Not. Roy. Astron. Soc.
  \textbf{318}, 625 (2000)

\bibitem{Kaiser_detect}
N.~Kaiser, G.~Wilson, G.A. Luppino, astro-ph/0003338  (2000)

\bibitem{vW_detect}
L.~van Waerbeke, et~al., Astron. Astrophys. \textbf{358}, 30 (2000)

\bibitem{Wittman_detect}
D.M. Wittman, J.A. Tyson, D.~Kirkman, I.~Dell'Antonio, G.~Bernstein, Nature
  \textbf{405}, 143 (2000)

\bibitem{FriTurHut}
J.~Frieman, M.~Turner, D.~Huterer, Ann. Rev. Astron. Astrophys. \textbf{46},
  385 (2008)

\bibitem{Mellier99}
Y.~Mellier, astro-ph/9901116  (1999)

\bibitem{Bartelmann_Schneider}
M.~Bartelmann, P.~Schneider, Phys. Rept. \textbf{340}, 291 (2001)

\bibitem{Refregier_ARAA}
A.~Refregier, Ann. Rev. Astron. Astrophys. \textbf{41}, 645 (2003)

\bibitem{Munshi_review}
D.~Munshi, P.~Valageas, L.~Van~Waerbeke, A.~Heavens, Phys. Rept. \textbf{462},
  67 (2008)

\bibitem{Hoekstra_Jain}
H.~{Hoekstra}, B.~{Jain}, Annual Review of Nuclear and Particle Science
  \textbf{58}, 99 (2008)

\bibitem{Hu_White}
W.~Hu, M.J. White, Astrophys. J. \textbf{554}, 67 (2001)

\bibitem{Shapiro_reduced}
C.~Shapiro, Astrophys. J. \textbf{696}, 775 (2009)

\bibitem{STEP1}
C.~Heymans, et~al., Mon. Not. Roy. Astron. Soc. \textbf{368}, 1323 (2006)

\bibitem{STEP2}
R.~Massey, et~al., Mon. Not. Roy. Astron. Soc. \textbf{376}, 13 (2007)

\bibitem{Hu_tomo}
W.~Hu, Astrophys. J. \textbf{522}, L21 (1999)

\bibitem{LSST}
P.~Abell, et~al., 0912.0201  (2009)

\bibitem{Cooray_Hu_bisp}
A.~Cooray, W.~Hu, Astrophys. J. \textbf{548}, 7 (2001)

\bibitem{Schneider_Lombardi_I}
P.~Schneider, M.~Lombardi, Astron. Astrophys. \textbf{397}, 809 (2003)

\bibitem{Schneider_Lombardi_II}
P.~Schneider, M.~Kilbinger, M.~Lombardi, Astron. Astrophys. \textbf{431}, 9
  (2005)

\bibitem{TJ_3pt_WL}
M.~Takada, B.~Jain, Mon. Not. Roy. Astron. Soc. \textbf{344}, 857 (2003)

\bibitem{Dodelson_Zhang_bisp}
S.~{Dodelson}, P.~{Zhang}, Phys. Rev. D \textbf{72}(8), 083001 (2005)

\bibitem{Bernardeau_VIRMOS_3pt}
F.~Bernardeau, Y.~Mellier, L.~van Waerbeke, Astron. Astrophys. \textbf{389},
  l28 (2002)

\bibitem{Pen_VIRMOS_skew}
U.L. Pen, et~al., Astrophys. J. \textbf{592}, 664 (2003)

\bibitem{Semboloni_COSMOS_3pt}
E.~Semboloni, et~al., 1005.4941  (2010)

\bibitem{TJ_PS+Bisp}
M.~Takada, B.~Jain, Mon. Not. Roy. Astron. Soc. \textbf{348}, 897 (2004)

\bibitem{Takada_Bridle}
M.~Takada, S.~Bridle, New J. Phys. \textbf{9}, 446 (2007)

\bibitem{Fischer}
P.~Fischer, et~al., Astron. J. \textbf{120}, 1198 (2000)

\bibitem{McKay}
T.A. McKay, et~al., astro-ph/0108013  (2001)

\bibitem{Hoek_Yee_Glad}
H.~Hoekstra, H.K.C. Yee, M.D. Gladders, Astrophys. J. \textbf{606}, 67 (2004)

\bibitem{Mandelbaum_starform}
R.~Mandelbaum, U.~Seljak, G.~Kauffmann, C.M. Hirata, J.~Brinkmann, Mon. Not.
  Roy. Astron. Soc. \textbf{368}, 715 (2006)

\bibitem{Wilson:2001}
G.~{Wilson}, N.~{Kaiser}, G.A. {Luppino}, L.L. {Cowie}, Astrophys. J.
  \textbf{555}, 572 (2001)

\bibitem{Sheldon04}
E.S. Sheldon, et~al., Astron. J. \textbf{127}, 2544 (2004)

\bibitem{Johnston05}
D.E. Johnston, et~al., Astrophys. J. \textbf{656}, 27 (2007)

\bibitem{Guzik_Seljak}
J.~Guzik, U.~Seljak, Mon. Not. Roy. Astron. Soc. \textbf{335}, 311 (2002)

\bibitem{Leauthaud_measure}
A.~{Leauthaud}, et~al., Astrophys. J. Suppl. \textbf{172}, 219 (2007)

\bibitem{Leauthaud_inprep}
A.~Leauthaud, et~al., in preparation

\bibitem{Kleinheinrich04}
M.~{Kleinheinrich}, P.~{Schneider}, H.~{Rix}, T.~{Erben}, C.~{Wolf},
  M.~{Schirmer}, K.~{Meisenheimer}, A.~{Borch}, S.~{Dye}, Z.~{Kovacs},
  L.~{Wisotzki}, Astron. Astrophys. \textbf{455}, 441 (2006)

\bibitem{Mandelbaum_profiles}
R.~Mandelbaum, et~al., Mon. Not. Roy. Astron. Soc. \textbf{372}, 758 (2006)

\bibitem{Johnston07}
D.E. Johnston, et~al., arXiv:0709.1159  (2007)

\bibitem{Leauthaud_ML}
A.~Leauthaud, et~al., Astrophys. J. \textbf{709}, 97 (2010)

\bibitem{Sheldon09_ML}
E.S. Sheldon, et~al., Astrophys. J. \textbf{703}, 2232 (2009)

\bibitem{Hu_Jain}
W.~Hu, B.~Jain, Phys. Rev. \textbf{D70}, 043009 (2004)

\bibitem{Schmidt_MGWL}
F.~Schmidt, Phys. Rev. \textbf{D78}, 043002 (2008)

\bibitem{Wittman01}
D.~Wittman, J.A. Tyson, V.E. Margoniner, J.G. Cohen, I.P. Dell'Antonio,
  Astrophys. J. \textbf{557}, L89 (2001)

\bibitem{Wittman02}
D.~Wittman, V.E. Margoniner, J.A. Tyson, J.G. Cohen, I.P. Dell'Antonio,
  Astrophys. J. \textbf{597}, 218 (2003)

\bibitem{Wittman06}
D.~Wittman, et~al., Astrophys. J. \textbf{643}, 128 (2006)

\bibitem{Schirmer07}
M.~{Schirmer}, T.~{Erben}, M.~{Hetterscheidt}, P.~{Schneider}, Astron.
  Astrophys. \textbf{462}, 875 (2007)

\bibitem{Dietrich07}
J.P. {Dietrich}, T.~{Erben}, G.~{Lamer}, P.~{Schneider}, A.~{Schwope},
  J.~{Hartlap}, M.~{Maturi}, Astron. Astrophys. \textbf{470}, 821 (2007)

\bibitem{Miyazaki07}
S.~{Miyazaki}, T.~{Hamana}, R.S. {Ellis}, N.~{Kashikawa}, R.J. {Massey},
  J.~{Taylor}, A.~{Refregier}, Astrophys. J. \textbf{669}, 714 (2007)

\bibitem{Clowe_Bullet}
D.~Clowe, et~al., Astrophys. J. \textbf{648}, L109 (2006)

\bibitem{Marian_Bernstein}
L.~Marian, G.M. Bernstein, Phys. Rev. \textbf{D73}, 123525 (2006)

\bibitem{Fang_Haiman}
W.J. Fang, Z.~Haiman, Phys. Rev. \textbf{D75}, 043010 (2007)

\bibitem{Hu_Keeton}
W.~Hu, C.R. Keeton, Phys. Rev. \textbf{D66}, 063506 (2002)

\bibitem{White_vW_Mackey}
M.J. White, L.~van Waerbeke, J.~Mackey, Astrophys. J. \textbf{575}, 640 (2002)

\bibitem{Padm_Seljak_Pen}
N.~Padmanabhan, U.~Seljak, U.L. Pen, New Astron. \textbf{8}, 581 (2003)

\bibitem{Hamana_Takada_Yoshida}
T.~Hamana, M.~Takada, N.~Yoshida, Mon. Not. Roy. Astron. Soc. \textbf{350}, 893
  (2004)

\bibitem{Hennawi_Spergel}
J.F. {Hennawi}, D.N. {Spergel}, Astrophys. J. \textbf{624}, 59 (2005)

\bibitem{Marian_Smith_Bernstein}
L.~Marian, R.E. Smith, G.M. Bernstein, Astrophys. J. \textbf{698}, L33 (2009)

\bibitem{Dietrich_Hartlap}
J.P. Dietrich, J.~Hartlap, Mon. Not. Roy. Astron. Soc. \textbf{in press} (2010)

\bibitem{Kratochvil}
J.M. Kratochvil, Z.~Haiman, M.~May, Phys. Rev. \textbf{D81}, 043519 (2010)

\bibitem{Dahle06}
H.~Dahle, Astrophys. J. \textbf{653}, 954 (2006)

\bibitem{Rozo09}
E.~{Rozo}, et~al., Astrophys. J. \textbf{708}, 645 (2010)

\bibitem{Jarvis_CTIO}
M.~{Jarvis}, G.M. {Bernstein}, P.~{Fischer}, D.~{Smith}, B.~{Jain}, J.A.
  {Tyson}, D.~{Wittman}, Astron. J. \textbf{125}, 1014 (2003)

\bibitem{Brown_COMBO17}
M.L. Brown, et~al., Mon. Not. Roy. Astron. Soc. \textbf{341}, 100 (2003)

\bibitem{Hoekstra_RCS}
H.~{Hoekstra}, H.K.C. {Yee}, M.D. {Gladders}, L.F. {Barrientos}, P.B. {Hall},
  L.~{Infante}, Astrophys. J. \textbf{572}, 55 (2002)

\bibitem{vanWaerbeke_VIRMOS}
L.~{Van Waerbeke}, Y.~{Mellier}, H.~{Hoekstra}, Astron. Astrophys.
  \textbf{429}, 75 (2005)

\bibitem{Hoekstra_CFHT}
H.~{Hoekstra}, Y.~{Mellier}, L.~{van Waerbeke}, E.~{Semboloni}, L.~{Fu}, M.J.
  {Hudson}, L.C. {Parker}, I.~{Tereno}, K.~{Benabed}, Astrophys. J.
  \textbf{647}, 116 (2006)

\bibitem{Hetterscheidt_GaBoDS}
M.~{Hetterscheidt}, P.~{Simon}, M.~{Schirmer}, H.~{Hildebrandt},
  T.~{Schrabback}, T.~{Erben}, P.~{Schneider}, "Astron. Astrophys.
  \textbf{468}, 859 (2007)

\bibitem{Benjamin}
J.~Benjamin, et~al., Mon. Not. Roy. Astron. Soc. \textbf{381}, 702 (2007)

\bibitem{Refregier_HST_MDS}
A.~Refregier, J.~Rhodes, E.J. Groth, Astrophys. J. \textbf{572}, L131 (2002)

\bibitem{Rhodes_STIS}
J.~Rhodes, et~al., Astrophys. J. \textbf{605}, 29 (2004)

\bibitem{Massey_Nature}
R.~Massey, et~al., Nature \textbf{445}, 286 (2007)

\bibitem{Kilbinger_CFHTLS}
M.~Kilbinger, et~al., Astron. Astrophys. \textbf{497}, 677 (2009)

\bibitem{Massey_COSMOS}
R.~Massey, et~al., Astrophys. J. Suppl. \textbf{172}, 239 (2007)

\bibitem{Fu_CFHTLS}
L.~Fu, et~al., Astron. Astrophys. \textbf{479}, 9 (2008)

\bibitem{WMAP3}
D.N. {Spergel}, et~al., Astrophys. J. Suppl. \textbf{170}, 377 (2007)

\bibitem{Jarvis_CTIO_DE}
M.~Jarvis, B.~Jain, G.~Bernstein, D.~Dolney, Astrophys. J. \textbf{644}, 71
  (2006)

\bibitem{Schimd_current_Q}
C.~Schimd, et~al., Astron. Astrophys. \textbf{463}, 405 (2007)

\bibitem{Hu_Tegmark_WL}
W.~Hu, M.~Tegmark, Astrophys. J. \textbf{514}, L65 (1999)

\bibitem{Huterer_thesis}
D.~Huterer, Phys. Rev. \textbf{D65}, 063001 (2002)

\bibitem{Simon_King_Schneider}
P.~Simon, L.J. King, P.~Schneider, Astron. Astrophys. \textbf{417}, 873 (2004)

\bibitem{Munshi_Valageas}
D.~Munshi, P.~Valageas, astro-ph/0510266  (2005)

\bibitem{Heavens_Kitching_Taylor}
A.F. Heavens, T.D. Kitching, A.N. Taylor, Mon. Not. Roy. Astron. Soc.
  \textbf{373}, 105 (2006)

\bibitem{Amara_Refregier}
A.~Amara, A.~Refregier, Mon. Not. Roy. Astron. Soc. \textbf{381}, 1018 (2007)

\bibitem{Zhan_Knox}
H.~Zhan, L.~Knox, astro-ph/0611159  (2006)

\bibitem{Shapiro_Dodelson_WL_counts}
C.~Shapiro, S.~Dodelson, Phys. Rev. \textbf{D76}, 083515 (2007)

\bibitem{Bernstein_comprehensive}
G.M. Bernstein, Astrophys. J. \textbf{695}, 652 (2009)

\bibitem{SNAP}
G.~Aldering, et~al., astro-ph/0405232  (2004)

\bibitem{Bernstein_Jain_cosmography}
G.M. Bernstein, B.~Jain, Astrophys. J. \textbf{600}, 17 (2004)

\bibitem{Jain_Taylor}
B.~Jain, A.~Taylor, Phys. Rev. Lett. \textbf{91}, 141302 (2003)

\bibitem{Zhang_Hui_Stebbins}
J.~Zhang, L.~Hui, A.~Stebbins, Astrophys. J. \textbf{635}, 806 (2005)

\bibitem{Bernstein_metric}
G.~Bernstein, Astrophys. J. \textbf{637}, 598 (2006)

\bibitem{Taylor_shear_ratio}
A.N. Taylor, T.D. Kitching, D.J. Bacon, A.F. Heavens, Mon. Not. Roy. Astron.
  Soc. \textbf{374}, 1377 (2007)

\bibitem{MMG}
D.~Huterer, E.V. Linder, Phys. Rev. \textbf{D75}, 023519 (2007)

\bibitem{Uzan_Bernardeau}
J.P. Uzan, F.~Bernardeau, Phys. Rev. \textbf{D64}, 083004 (2001)

\bibitem{Jain_Zhang}
B.~Jain, P.~Zhang, Phys. Rev. \textbf{D78}, 063503 (2008)

\bibitem{Uzan_GRG}
J.P. Uzan, arXiv:0908.2243  (2009)

\bibitem{Linder_gamma}
E.V. {Linder}, Phys. Rev. \textbf{D72}(4), 043529 (2005)

\bibitem{LinCah07}
E.V. {Linder}, R.N. {Cahn}, Astroparticle Physics \textbf{28}, 481 (2007)

\bibitem{DGP}
G.R. Dvali, G.~Gabadadze, M.~Porrati, Phys. Lett. \textbf{B485}, 208 (2000)

\bibitem{IshUpaSpe}
M.~Ishak, A.~Upadhye, D.N. Spergel, Phys. Rev. \textbf{D74}, 043513 (2006)

\bibitem{Daniel}
S.F. Daniel, R.R. Caldwell, A.~Cooray, A.~Melchiorri, Phys. Rev. D
  \textbf{77}(10), 103513 (2008)

\bibitem{Amend_Kunz_Sapone}
L.~{Amendola}, M.~{Kunz}, D.~{Sapone}, Journal of Cosmology and Astro-Particle
  Physics \textbf{4}, 13 (2008)

\bibitem{Dore_CFHTLS}
O.~{Dor{\'e}}, et~al., arXiv:0712.1599  (2007)

\bibitem{White_Kochanek}
M.J. White, C.S. Kochanek, Astrophys. J. \textbf{560}, 539 (2001)

\bibitem{Reyes:2010tr}
R.~Reyes, et~al., Nature \textbf{464}, 256 (2010)

\bibitem{Schmidt08}
F.~Schmidt, Phys. Rev. \textbf{D78}, 043002 (2008)

\bibitem{Zhan_Knox_Tyson}
H.~Zhan, L.~Knox, J.A. Tyson, Astrophys. J. \textbf{690}, 923 (2009)

\bibitem{Zhang_Bean_etal}
P.~Zhang, R.~Bean, M.~Liguori, S.~Dodelson, arXiv:0809.2836  (2008)

\bibitem{Zhao_etal_08}
G.B. Zhao, L.~Pogosian, A.~Silvestri, J.~Zylberberg, Phys. Rev. \textbf{D79},
  083513 (2009)

\bibitem{CooHolHut_SNmag}
A.~Cooray, D.~Holz, D.~Huterer, Astrophys. J. \textbf{637}, L77 (2006)

\bibitem{Dodelson_Vallinotto}
S.~Dodelson, A.~Vallinotto, Phys. Rev. \textbf{D74}, 063515 (2006)

\bibitem{MaHuHut}
Z.M. Ma, W.~Hu, D.~Huterer, Astrophys. J. \textbf{636}, 21 (2005)

\bibitem{Oyaizu07}
H.~Oyaizu, M.~Lima, C.E. Cunha, H.~Lin, J.~Frieman, Astrophys. J. \textbf{689},
  709 (2008)

\bibitem{catastrophic}
G.~Bernstein, D.~Huterer, Mon. Not. Roy. Astron. Soc. \textbf{401}, 1399 (2010)

\bibitem{Hearin_cata}
A.P. Hearin, A.R. Zentner, Z.~Ma, D.~Huterer, arXiv:1002.3383  (2010)

\bibitem{Bernstein_Ma}
Z.~Ma, G.~Bernstein, Astrophys. J. \textbf{682}, 39 (2008)

\bibitem{HTBJ}
D.~Huterer, M.~Takada, G.~Bernstein, B.~Jain, Mon. Not. Roy. Astron. Soc.
  \textbf{366}, 101 (2006)

\bibitem{Kitching_sys}
T.D. {Kitching}, A.N. {Taylor}, A.F. {Heavens}, Mon. Not. Roy. Astron. Soc.
  \textbf{389}, 173 (2008)

\bibitem{Zhang_selfcal}
P.~Zhang, arXiv:0811.0613  (2008)

\bibitem{Zhang_Pen_Bernstein}
P.~Zhang, U.L. Pen, G.~Bernstein, arXiv:0910.4181  (2009)

\bibitem{Huterer_Takada}
D.~Huterer, M.~Takada, Astropart. Phys. \textbf{23}, 369 (2005)

\bibitem{Heitmann_robustness}
K.~Heitmann, P.M. Ricker, M.S. Warren, S.~Habib, Astrophys. J. Suppl.
  \textbf{160}, 28 (2005)

\bibitem{Rudd_Zentner_Kravtsov}
D.H. Rudd, A.R. Zentner, A.V. Kravtsov, Astrophys. J. \textbf{672}, 19 (2008)

\bibitem{Zentner_Rudd_Hu}
A.R. Zentner, D.H. Rudd, W.~Hu, Phys. Rev. \textbf{D77}, 043507 (2008)

\bibitem{Hearin_Zentner}
A.P. {Hearin}, A.R. {Zentner}, Journal of Cosmology and Astro-Particle Physics
  \textbf{4}, 32 (2009)

\bibitem{Heitmann_calib}
K.~Heitmann, D.~Higdon, C.~Nakhleh, S.~Habib, Astrophys. J. \textbf{646}, L1
  (2006)

\bibitem{Heitmann_CoyoteII}
K.~Heitmann, et~al., Astrophys. J. \textbf{705}, 156 (2009)

\bibitem{Takada_White}
M.~Takada, M.J. White, Astrophys. J. \textbf{601}, L1 (2004)

\bibitem{Joachimi_Schneider_nulling}
B.~Joachimi, P.~Schneider, Astron. Astrophys. \textbf{488}, 829 (2008)

\bibitem{nulling}
D.~Huterer, M.J. White, Phys. Rev. \textbf{D72}, 043002 (2005)

\bibitem{Hanson_GRG}
D.~Hanson, A.~Challinor, A.~Lewis, arXiv:0911.0612  (2009)

\end{thebibliography}

\end{document}